\documentclass[aps,prd,twocolumn,showpacs,preprintnumbers,nofootinbib]{revtex4}

\usepackage{graphicx}
\usepackage[FIGTOPCAP]{subfigure}
\usepackage{dcolumn}
\usepackage{bm}
\usepackage[normalem]{ulem}

\usepackage{amsmath}
\usepackage{epstopdf}
\usepackage{amsmath,amssymb}

\newcommand{\mbf}[1]{\ensuremath{\boldsymbol{#1}}}

\begin{document}

\title{Heavy quarkonia in strong magnetic fields}

\author{Claudio Bonati}
\email{claudio.bonati@df.unipi.it}
\affiliation{INFN - Sezione di Pisa, Largo Pontecorvo 3, I-56127 Pisa, Italy}

\author{Massimo D'Elia}
\email{massimo.delia@unipi.it}
\affiliation{INFN - Sezione di Pisa, Largo Pontecorvo 3, I-56127 Pisa, Italy}
\affiliation{Dipartimento di Fisica dell'Universit\`a di Pisa, Largo Pontecorvo 3, I-56127 Pisa, Italy}

\author{Andrea Rucci}
\email{andrea.rucci@pi.infn.it}
\affiliation{Dipartimento di Fisica dell'Universit\`a di Pisa, Largo Pontecorvo 3, I-56127 Pisa, Italy}

\date{\today}

\begin{abstract}
We investigate the influence of a homogeneus and constant strong external
magnetic field on the heavy-meson spectrum. Quarkonium states $c\bar{c}$ and
$b\bar{b}$ are described within a non-relativistic framework and by means of a
suitable potential model based on the Cornell parametrization. In particular,
in this work we propose a model which takes into account the possible
anisotropies emerging at the level of the static quark-antiquark potential, as
observed in recent lattice studies.  The investigation is perfomed both with
and without taking into account the anisotropy of the static potential, in
order to better clarify its effects.
\end{abstract}


\maketitle

\section{Introduction}\label{sec:intro}

In recent times there has been a great interest regarding the physics of
strongly interacting matter in the presence of strong external magnetic fields,
i.e.  such that\footnote{For reference, a magnetic field of the order of
$10^{15}$ Tesla corresponds to $eB \simeq 3.3\, m_{\pi}^2\approx
0.06\,\mathrm{GeV}^2$.} $eB\simeq m_{\pi}^2$ or larger (see, e.g.,
Refs.~\cite{lecnotmag, misho} for recent reviews).  This topic could be
relevant to the study of some dense astrophysical objects, like
magnetars~\cite{magnetars}, and for cosmology~\cite{vacha, grarub}. However,
the main interest was triggered by the fact that magnetic fields of this order
of magnitude can be created in a laboratory, in particular in heavy-ion
collisions~\cite{hi1, hi2, hi3, hi4, tuchin}, when two relativistic heavy ions
collide with a non-zero impact parameter, producing a huge field in the
collision region.  For example, one can reach $|e|B\simeq0.2-0.3~\rm{GeV}^2$ in
Pb+Pb collisions at the Large Hadron Collider (LHC).

Such huge magnetic fields are produced in the very early stages of the
collision.  It is still not clear how long and to what extent they survive the
thermalization process of the fireball created after the collision.  Therefore,
while various theoretical investigations, based both on model studies and on
Lattice QCD simulations (LQCD), have predicted many interesting phenomena
affecting the properties of strongly interacting matter in the presence of
strong magnetic backgrounds, it is still uncertain to what extent such
phenomena will be detectable in heavy ion experiments.

In this perspective, effects regarding the physics of heavy flavors are of
particular interest, since they are more sensitive to the conditions taking
place in the early stages of the collision.  Various studies have approached
the issue of  quarkonia spectra and production rates in the presence of
magnetic backgrounds~\cite{Machado1, Machado2, Alford, morita, dudalmertens,
Filip, Xu}.  Many interesting phenomena have been predicted, including the
emergence of magnetic field induced mixings between different states and of
production anisotropies with respect to the collision plane.

The starting point for most of these investigations is the coupling of the
magnetic field with electric charges and magnetic moments carried by the
valence quarks. However, the magnetic field is known to induce important
modifications also at a non-perturbative level and in the gluon sector: a
natural question is whether that can change the picture both quantitatively and
qualitatively.

A very interesting phenomenon in this respect is the magnetic field induced
modification of the static quark-antiquark ($Q\bar{Q}$) potential. This is a
typical pure gauge quantity (it is related to the expectation values of Wilson
loops) which represents the starting point for many approaches to the study of
heavy quarkonia, tipically within a non-relativistic approximation. 
The static potential is usually expressed in terms of the so called Cornell
parametrization~\cite{Eichten:1974af}:
\begin{equation}\label{eq:cornell} 
V(r)=-\frac{\alpha}{r}+\sigma r\ .
\end{equation} 
where $\alpha$ is the Coulomb coupling and $\sigma$ the string tension.  

Such a non-relativistic approach is reliable only for heavy quark bound states,
for which the interaction energy is not a large fraction of the total mass, and for
not too large magnetic fields. As a rule of thumb one can use $eB\hbar/(m^2c^3)
\ll 1$ (where $m$ is the heavy quark mass) to estimate which magnetic fields
can be explored in this approach. In our worst case (charm quark and
$|e|B=0.3~\mathrm{GeV}^2$) the previous ratio is about $0.18$, a numerical
value which is roughly equal to the ratio $(M_{J/\Psi}-2m_c)/M_{J/\Psi}$,
which gives an estimate of the validity of the non-relativistic
approach also for $B = 0$.
For bottomonia states $eB\hbar/(m^2c^3)$ gets smaller by more than one 
order of
magnitude.

The magnetic background field breaks rotational invariance, hence one may
expect the emergence of anisotropies in the potential, which is central
otherwise.  The issue is clearly related to quark loop effects, since gluons
are not directly coupled to the magnetic field, and has been investigated
within various model studies~\cite{MS,Andreichikov:2012xe,
chernostring,Rougemont:2014efa,ferrer, taya, simonaniso}: the Coulomb coupling
has been predicted to change in the transverse
direction~\cite{MS,Rougemont:2014efa}, in the longitudinal
direction~\cite{Andreichikov:2012xe,ferrer} or in both~\cite{simonaniso};
regarding the string tension, while string theory studies do not predict an
influence of the magnetic field on it~\cite{Ferrer:1990na, Ferrara:1993sq,
Ferrer:1994pw, Ferrer:1995kc, Ambjorn:2000yr}, other approaches do (see, e.g.,
Refs.~\cite{chernostring,simonaniso}).

Further insight into the question has been provided by a recent lattice QCD
investigation~\cite{anisostring}: the potential gets steeper in the directions
transverse to the magnetic field and flatter in the longitudinal one.  In
particular one observes a larger (smaller) string tension in the transverse
(longitudinal) direction, while the opposite happens for the Coulomb
coupling~\cite{anisostring}.

The purpose of this study is to reconsider the computation of the heavy
quarkonia spectrum and mixings in the presence of a magnetic field, in the
light of the existing anisotropy in the static quark-antiquark potential.
Since standard magnetic field effects decrease with the mass of the valence
quarks, while the static potential remains unchanged, we might expect that
corrections be relatively more important for bottomonia than for charmonia.
 
We will follow the standard non-relativistic two-body approach (see, e.g.,
Refs.~\cite{Bali:2000gf, Brambilla} for reviews on this subject), which can be
tackled using standard numerical methods.  In particular, our treatment will be
close to that of Ref.~\cite{Alford}, apart from the presence of the anisotropy,
which will be included according to the lattice results of
Ref.~\cite{anisostring}.  Since in Ref.~\cite{anisostring} the potential has
been studied only along two directions (transverse and longitudinal), we shall
adopt the simplest possible parametrization which takes into account generic
angles.

The paper is organized as follows.  In Sec.~\ref{sec:twobody} we will review
the non-relativistic approach to the two-body problem in an external magnetic
field. In Sec.~\ref{qqpot} we will discuss the form of the static potential in
a magnetic background and our parametrization for its angular dependence. In
Sec.~\ref{numres} we will discuss our numerical approach to the problem (more
details and results from a test on the harmonic oscillator are reported in the
Appendix) and present results for charmonia and bottomonia. Finally, in
Sec.~\ref{summary}, we will draw our conclusions.

\section{The quantum two-body problem in external magnetic field}\label{sec:twobody}

The non-relativistic Hamiltonian of two particles of masses $m_i$ and charges
$q_i$ ($i=1,2$) in an external magnetic field is:
\begin{equation}\label{eq:ham}
\begin{aligned}
\hat{H} = & \sum_{i=1}^2\frac{1}{2m_i}\left[\hat{\mbf{p}}_i-q_i\mbf{A}(\mbf{x}_i)\right]^2 + \\ 
& +V\left(\mbf{x}_1,\mbf{x}_2\right)-(\mbf{\mu}_1+\mbf{\mu}_2)\cdot\mbf{B}\ ,
\end{aligned}
\end{equation}
where $\mbf{\mu}_1$, $\mbf{\mu}_2$ are the magnetic moments of the two
particles. The presence of a vector potential $\mbf{A}(\mbf{x})$ makes the
system not invariant under translations, i.e.
$\mbf{x}_i\to\mbf{x}_i+\mbf{\alpha}$. Even in the case of an uniform magnetic
field, both the canonical and the kinetic momentum
$\hat{\mbf{P}}=\sum_{i=1}^2(\hat{\mbf{p}}_i-q_i\mbf{A}(\mbf{x}_i)\,)$ do not
commute with the Hamiltonian in Eq.~\eqref{eq:ham}.

An invariance group of $\hat{H}$ is obtained by simultaneously performing a
coordinate translation and a gauge transformation, i.e. by changing both
$\mbf{x}_i$ and $\mbf{p}_i$. The generator of this transformation is the
pseudomomentum operator (see \cite{Avron} for more details) and can be written
in a particularly simple form in the symmetric gauge\footnote{See \cite{Herold}
for the form of the pseudomomentum in a generic gauge.}
$\mbf{A}(\mbf{x})=\frac{1}{2}\mbf{B}\times\mbf{x}$, where it is given by:
\begin{equation}\label{eq:pseudomom}
\hat{\mbf{K}}=\sum_{i=1}^2\left( \hat{\mbf{p}}_i+\frac{1}{2}q_i\mbf{B}\times\mbf{x}_i\right)\ .
\end{equation}
From this expression, it is not obvious that different components of the
pseudomomentum are commuting observables, and in fact this is not in general
true. This is however what happens for globally neutral system (in particular for
the $Q\bar{Q}$ system to be studied in this paper), since the following
commutation relations hold
\begin{equation}
[\hat{K}_j, \hat{K}_{\ell}]=-i(q_1+q_2)\epsilon_{j\ell k}B_k\ .
\end{equation}

From now on we will explicitly restrict to the case of a particle-antiparticle
system and we will adopt the notation $q\equiv q_1=-q_2$ for the electric
charge and $m$ for the particle masses (see e.g. Ref.~\cite{Alford} for the
general case).  Written in terms of the relative coordinate
$\mbf{r}=\mbf{x}_1-\mbf{x}_2$, the pseudomomentum takes the form
\begin{equation}\label{eq:K}
\hat{\mbf{K}}=\hat{\mbf{p}}_1+\hat{\mbf{p}}_2+\frac{1}{2}q\mbf{B}\times\mbf{r}
=\hat{\mbf{P}}+q\mbf{B}\times\mbf{r}\ ,
\end{equation}
From this expression it appears natural to adopt the following \emph{ansatz} for
the eigenstate $\Phi$ of the two body Hamiltonian:
\begin{equation}\label{eq:ansatz}
\Phi(\mbf{R}, \mbf{r}, \mbf{\sigma})=\exp\left[i\left(\mbf{K}-\frac{1}{2}q\mbf{B}\times\mbf{r}
\right)\cdot \mbf{R}\right]\Psi(\mbf{r},\mbf{\sigma})\ ,
\end{equation}
where $\mbf{K}$ denotes the eigenvalue of the operator $\hat{\mbf{K}}$, $\mbf{R}$ is 
the coordinate of the center of mass and $\mbf{\sigma}$ is a shorthand for the spin variables. Using
the expression Eq.~\eqref{eq:ansatz}, the Hamiltonian in Eq.~\eqref{eq:ham} can
be rewritten in the form (acting on the reduced wave function $\Psi$):
\begin{equation}\label{eq:redham}
\begin{aligned}
\hat{H}=&\frac{\mbf{K}^2}{2M}-\frac{q}{M}(\mbf{K}\times \mbf{B})\cdot\mbf{r} -\frac{\nabla^2}{2\mu}+\\
&+\frac{q^2}{2\mu}(\mbf{B}\times\mbf{r})^2+V(\mbf{r})-(\mbf{\mu}_1+\mbf{\mu}_2)\cdot\mbf{B}\ ,
\end{aligned}
\end{equation}
where $M\equiv 2m$ and $\mu\equiv m/2$ are respectively the total and the
reduced mass of the two body system. Because of the lack of translation
invariance, in this Hamiltonian the center of mass and the relative motion are
not decoupled, unless $\mbf{K}\times \mbf{B}=0$.

Given this nontrivial dependence on the $\mbf{K}$ value, one needs a
prescription to subtract the energy associated with the center of mass motion.
We will follow the idea presented in Ref.~\cite{Alford}: given the eigenvalue
$E$ of Eq.~\eqref{eq:redham}, we will consider $E-\frac{\langle
\hat{\mbf{P}}\rangle^2}{2M}$ as the corrected binding energy, with the kinetic
momentum expectation value being computed by using Eq.~\eqref{eq:K}.

Finally, let us discuss the role of the spin interaction term in
Eq.~\eqref{eq:redham}. Quark magnetic moments may be expressed as
$\mbf{\mu}_i=g\mu_i\mbf{s}_i$, where $g=2$ is the quark $g$-factor in the
non-relativistic approximation, $\mu_i=q_i/2m_i$ is the quark magneton and
$\mbf{s}_i=\mbf{\sigma}^i/2$ is its spin. Therefore, in the case of a
quark-antiquark pair
\begin{equation}\label{eq:magnmom}
-\left(\mbf{\mu}_1+\mbf{\mu}_2\right)\cdot\mbf{B}=
-\frac{gq}{4m}\left(\mbf{\sigma}^1-\mbf{\sigma}^2\right)\cdot\mbf{B}\ .
\end{equation}
It can be verified that this term induces a mixing between the singlet
$|00\rangle$ and the triplet $|10\rangle$ spin states. Indeed, in the presence
of a magnetic field $\mbf{B}=B\mbf{\hat{z}}$, the operator in
Eq.~\eqref{eq:magnmom} has the only non-zero matrix element between spin states
given by 
\begin{equation}\label{eq:magnmomelem}
\langle00|\left(\mbf{\mu}_1+\mbf{\mu}_2\right)\cdot\mbf{B}|10\rangle =
-\frac{gqB}{2m} \ .  
\end{equation}
Such a term increases the energy of the triplet state and decreases that of the
singlet. When Eq.~\eqref{eq:magnmomelem} is used 
for a noncentral potential, or when
a spin-spin interaction is also present, some caution is needed, since
possible mixings between different orbital states can be present (see
Appendix.~\ref{sec:algorithm} for more details).

\section{The $Q\bar{Q}$ potential}
\label{qqpot}

It is well known that the main features of the quarkonium spectrum can be
understood by using a central potential between the heavy quarks of the so
called Cornell form reported in Eq.~\eqref{eq:cornell}.
This parametrization was introduced in Ref.~\cite{Eichten:1974af} on
phenomenological bases, as the simplest form of the potential that takes into
account both the short distance perturbative contribution and color
confinement.  It was later realized that the form Eq.~\eqref{eq:cornell} of the
potential correctly describes the spin averaged potential between two static
(i.e. infinitely massive) quarks as estimated from the first principles of QCD
by its lattice formulation (see, e.g., Ref.~\cite{Bali:2000gf}).

To properly describe the fine structure of the quarkonium spectrum further spin
dependent terms have to be added to the potential. All such corrections to the
static quark potential can in principle be obtained by an expansion in the
inverse quark mass (see Ref.~\cite{Brambilla}), however their precise
functional form is generally not well known and difficult to extract from LQCD
computations.  Because of this lacking of precise theoretical information,
several different parametrization of these terms exist, whose coefficients are
fixed by comparing with experimental results.

\begin{table}
\centering
\setlength{\tabcolsep}{10pt}
  \begin{tabular}{| c | c | c | c |}
    \hline
    $Q\bar{Q}$ & State & Name & Mass [MeV] \\
    \hline
    $c\bar{c}$ & $1^1S_0$ & $\eta_c$ & $2980.3 \pm 1.2$\\
    $\cdot$  & $1^3S_1$ & $J$/$\psi$ & $3096.916 \pm 0.011$\\
    $\cdot$ & $1^3P_0$ & $\chi_{c0}$ & $3414.75 \pm 0.31$\\
    $\cdot$ & $1^3P_1$ & $\chi_{c1}$ & $3510.66 \pm 0.07$\\
    $\cdot$ & $1^3P_2$ & $\chi_{c2}$ & $3556.20 \pm 0.09$\\
    $\cdot$ & $1^1P_1$ & $h_c$ & $3525.38 \pm 0.11$\\
    $\cdot$ & $2^1S_0$ & $\eta_c(2S)$ & $3639.4 \pm 1.3$\\
    $\cdot$ & $2^3S_1$ & $\psi(2S)$ & $3686.109 \pm 0.02$\\
    $b\bar{b}$ & $1^1S_0$ & $\eta_b$ & $9390.9 \pm 2.8$\\
    $\cdot$ & $1^3S_1$ & $\Upsilon$ & $9460.30 \pm 0.26$\\
    $\cdot$ & $1^3P_0$ & $\chi_{b0}$ & $9859.44 \pm 0.74$\\
    $\cdot$ & $1^3P_1$ & $\chi_{b1}$ & $9892.78 \pm 0.57$\\
    $\cdot$ & $1^3P_2$ & $\chi_{b2}$ & $9912.21 \pm 0.57$\\
    $\cdot$ & $1^1P_1$ & $h_b$ & $9898.3 \pm 1.1$\\
    $\cdot$ & $2^1S_0$ & $\eta_b(2S)$ & $9999.0 \pm 3.5$\\
    $\cdot$ & $2^3S_1$ & $\Upsilon(2S)$ & $10232.26 \pm 0.31$\\
    \hline
  \end{tabular}
  \caption{Lowest part of charmonium and bottomonium mass spectrum. 
           Data from  \cite{pdg2012}.}
\label{table:mesonspectrum}
\end{table}

Since we are mainly interested in the lowest quarkonium levels (see
Tab.~\ref{table:mesonspectrum}), the most important contribution to be added to
the Cornell potential is the spin-spin interaction, responsible e.g., for the
mass splitting between the $J/\Psi$ and $\eta_c$ levels of charmonium.  We will
adopt the parametrization
\begin{equation}\label{eq:spinspin}
V_{\sigma\sigma}=(\mbf{\sigma}_1\cdot\mbf{\sigma}_2)\gamma e^{-\beta r}\ ,
\end{equation}
where $\beta$ and $\gamma$ are phenomenological constants, whose values are
reported in Tab.~\ref{table:potparam}. This form of the spin-spin interaction
is supported by the lattice result presented in Ref.~\cite{Kawanai} and was
previously used in Ref. \cite{Alford} to study the influence of a magnetic
field on the meson spectrum.

\begin{figure}[htb!]
\includegraphics[width=\columnwidth]{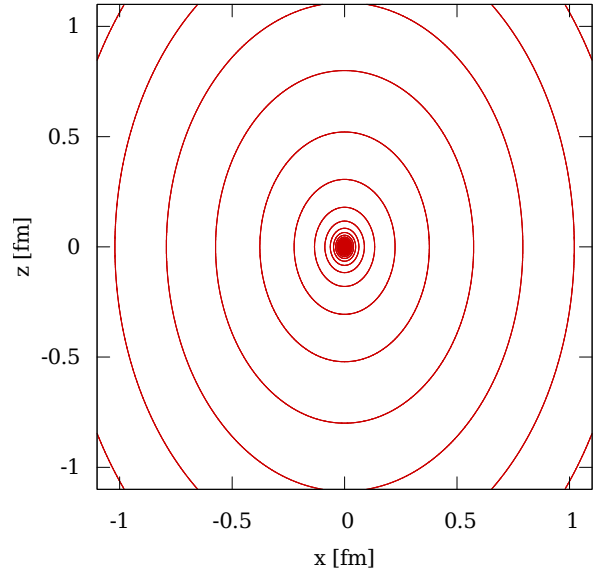}
\caption{Contour map of the potential in Eq.~\eqref{eq:potB} on the $x-z$ plane
for $B=0.6\,\mathrm{GeV}^2$, with $\mbf{B}$ directed along the $\mbf{\hat{z}}$ axis. 
Parameters used are the ones for charmonium (see Tab.~\ref{table:potparam}).
}
\label{fig:contourpot}
\end{figure}

While in previous studies the static Cornell potential in
Eq.~\eqref{eq:cornell} was assumed to be independent of the magnetic field, in
this work we will consider also the anisotropy observed in
Ref.~\cite{anisostring}. However, since in Ref.~\cite{anisostring} only the
values of the potential along the coordinate axes have been investigated, we
are forced to make an \emph{ansatz} on the form of the potential, with the
constraint that it reproduces the observed behaviour on the axes. To this aim,
we propose the following anisotropic form of the static potential terms
($\mbf{B}$ is directed along the $\mbf{\hat{z}}$ axis)
\begin{equation}
\begin{aligned}
\frac{\alpha}{r} &\to \frac{\alpha}{\sqrt{\epsilon_{xy}^{(\alpha)}(x^2+y^2)+\epsilon_{z}^{(\alpha)}z^2}}\\
\sigma r &\to \sigma\sqrt{\epsilon_{xy}^{(\sigma)}(x^2+y^2)+\epsilon_{z}^{(\sigma)}z^2}
\end{aligned}
\end{equation}
where the scaling parameters $\epsilon_{xy}(B)$ and $\epsilon_z(B)$ are
functions of the external magnetic field. Such a parametrization is inspired by
that of the electrostatic interaction in the presence of an anisotropic
dielectric constant (see Ref.~\cite{Landau} \S 13). It is possible to rewrite
the expression above in a more usual fashion, by absorbing the angular and $B$
dependences into the $\alpha$ and $\sigma$ parameters
\begin{equation}\label{eq:potB} 
V(r,\theta, B)=-\frac{\alpha(\theta, B)}{r}+\sigma(\theta,B) r\ ,
\end{equation}
where $\theta$ is the spherical azimuthal angle and
\begin{equation}
\begin{aligned}
\alpha(\theta,B)&=\frac{\alpha}{\epsilon_1^{(\alpha)}\sqrt{1+\epsilon_2^{(\alpha)}(B)\sin^2(\theta)}}\\
\sigma(\theta,B)&=\sigma\epsilon_1^{(\sigma)}(B)\sqrt{1+\epsilon_2^{(\sigma)}(B)\sin^2(\theta)}
\end{aligned}
\end{equation}
with $\epsilon_i^{(O)}$ related to the scaling parameters previously defined by
setting $\epsilon_1=\sqrt{\epsilon_z}$ and
$\epsilon_2=\epsilon_{xy}/\epsilon_z-1$. According to Ref. \cite{anisostring},
when increasing the magnetic field the string tension gets larger in the
transverse directions, while it decreases in the longitudinal $\hat{z}$
direction; the opposite behaviour is observed for the coefficient $\alpha$ of
the Coulomb term.  To reproduce such a behaviour, the coefficients
$\epsilon_i^{(O)}$ must be positive. In particular, they are related to the
coefficients $A, C$ estimated in Ref. \cite{anisostring} (see
Tab.~\ref{table:anisoparam}) by the relations
\begin{equation}
\begin{aligned}
& \epsilon_1^{(\alpha)}(B)=\left(1+A^{\alpha_{z}}(|e|B)^{C^{\alpha_{z}}}\right)^{-1} \\
& \epsilon_2^{(\alpha)}(B)=\left(\frac{1+A^{\alpha_{z}}(|e|B)^{C^{\alpha_{z}}}}{1+A^{\alpha_{xy}}(|e|B)^{C^{\alpha_{xy}}}}\right)^2-1 \\
& \epsilon_1^{(\sigma)}(B)=\left(1+A^{\sigma_{z}}(|e|B)^{C^{\sigma_{z}}}\right) \\
& \epsilon_2^{(\sigma)}(B)=\left(\frac{1+A^{\sigma_{xy}}(|e|B)^{C^{\sigma_{xy}}}}{1+A^{\sigma_{z}}(|e|B)^{C^{\sigma_{z}}}}\right)^2-1 \, .
\end{aligned}
\label{eq:anisoepsilon}
\end{equation}
A graphical representation of the anisotropic potential is shown in Fig.
\ref{fig:contourpot}.

\begin{table}
\begin{tabular}{ |c|c|c|c|c| } \hline $O$ & \rule{0mm}{3.5mm} $A^{O_{xy}}$ & $C^{O_{xy}}$ & $A^{O_{z}}$ &
$C^{O_{z}}$ \\ \hline $\sigma$ & 0.29$\pm$0.02 & 0.9$\pm$0.1 & -0.34$\pm$0.01 &
1.5$\pm$0.1 \\ $\alpha$ & -0.24$\pm$0.04 & 0.7$\pm$0.2 & 0.24$\pm$0.03 & 1.7$\pm$0.4 \\
\hline
\end{tabular}
\caption{Coefficients $A$, $C$ estimated in \cite{anisostring}.}
\label{table:anisoparam}
\end{table}

Finally, let us briefly discuss the role of the spin-spin interaction term
\eqref{eq:spinspin} within the anisotropic potential picture. In principle,
both parameters, $\beta$ and $\gamma$, may acquire a dependence on the external
magnetic field like that observed for the string tension and  the
Coulomb coupling. Such a dependence has not yet been investigated and no ansatz
is available. However, since this term represents a spin-dependent relativistic
correction of order $m^{-2}$~\cite{Eichten79,Eichten81, Bali:2000gf,
Brambilla}, it is reasonable to expect that also the $B$-dependence be of the
same order in the quark mass $m$, hence much weaker than that induced on the
spin-independent part of the potential.  Therefore in the following we will
make use of the same values for $\beta$ and $\gamma$ adopted in
Ref.~\cite{Alford}.  A similar attitude will be adopted with respect to other
relativistic corrections, like the spin-orbit term.

\section{Computation and numerical results}
\label{numres}

In this section we report our results for the dependence on the magnetic field
of the masses of the $1S$ and $1P$ states for both charmonium and bottomonium.  

To this purpose we used the Hamiltonian in Eq.~\eqref{eq:redham}, with a 
magnetic
field $\mbf{B}=B\mbf{\hat{z}}$ and pseudo-momentum $\mbf{K}=K\mbf{\hat{x}}$,
with the anisotropic potential parametrized as in Eq.~\eqref{eq:potB}, together
with the spin-spin interaction term Eq.~\eqref{eq:spinspin}.  Values of the
parameters adopted are reported in Tab.~\ref{table:potparam} and have been fixed
according to Ref.~\cite{Alford}, thus enabling a direct comparison with the
case in which the anisotropy of the potential is neglected.  Numerical values
of the anisotropy parameters, defined in Eq.~\eqref{eq:anisoepsilon}, have been
fixed by using the coefficients shown in Tab.~\ref{table:anisoparam}. To take
into account the uncertainty associated with these parameters, several
simulations were performed, corresponding to different combinations of these
values.

\begin{table}
\begin{tabular}{| c | c | c | c | c | c |}
  \hline & $\gamma$ & $\beta$ & $\alpha$ & $\sigma$ & $m$\\
  \hline 
  $c\bar{c}$ & 2.060 GeV & 1.982 GeV & 0.312 & \rule{0mm}{3.4mm} 0.174 GeV$^2$ & 1.29 GeV\\
  $b\bar{b}$ & 0.318 GeV & 1.982 GeV & 0.421 & \rule{0mm}{3.4mm} 0.210 GeV$^2$ & 4.70 GeV\\
  \hline
\end{tabular}
\caption{Parameters used for the potential (same as in Ref.~\cite{Alford}).}
\label{table:potparam}
\end{table}

In order to allow for a better physical comprehension of the results, the mass
spectrum will be shown as a function of the mean kinetic momentum
$\langle\mbf{P}\rangle$, instead of using the pseudo-momentum $\mbf{K}$. To
this purpose we solved the system for several values of $\mbf{K}$, computed the
value of the mean kinetic momentum for each eigenstate and interpolated these
values to the desired point.  Details of the numerical algorithm used to
extract eigenstates are reported in Appendix~\ref{sec:algorithm}, together with
a test of the algorithm in an analytically solvable case.

\subsection{Charmonium states}

\begin{figure}[htb!]
\subfigure[]{\includegraphics[width=\columnwidth]{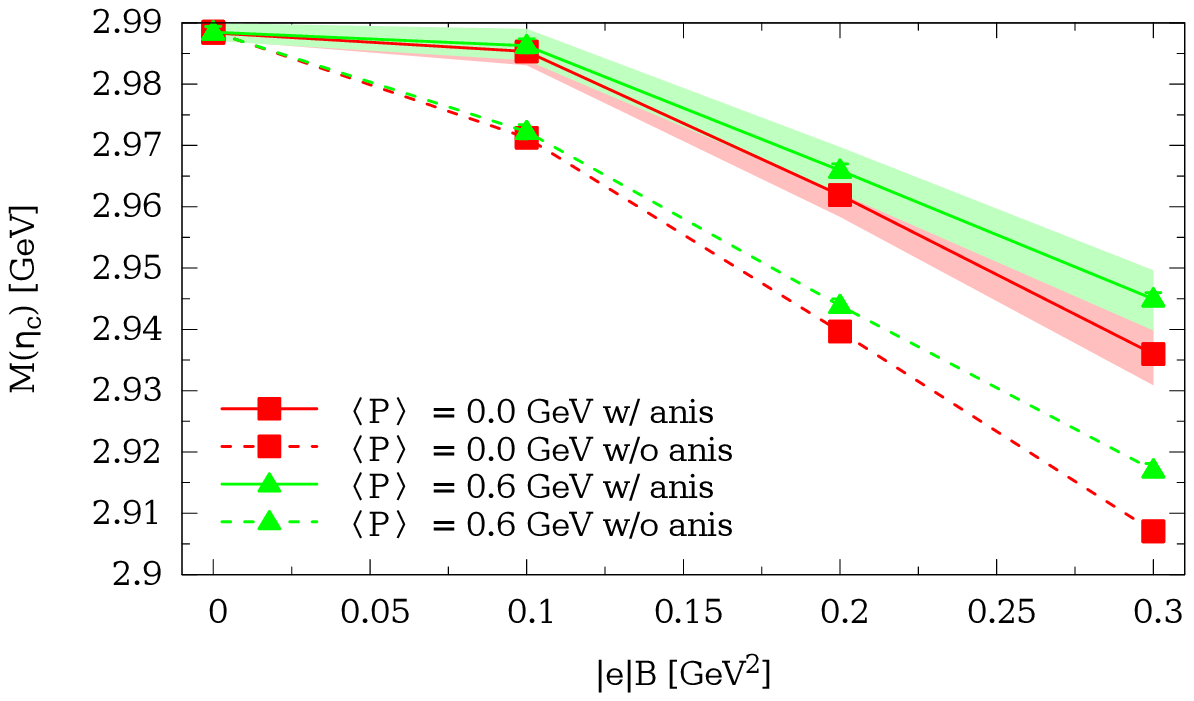}}
\subfigure[]{\includegraphics[width=\columnwidth]{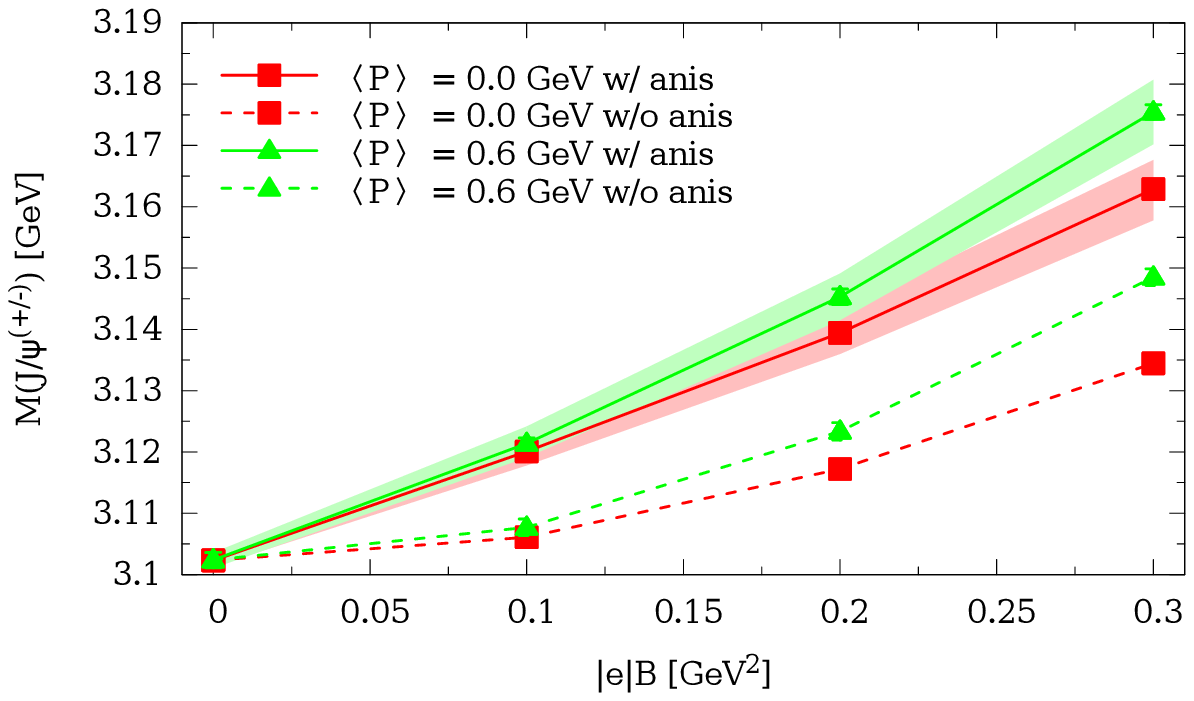}}
\subfigure[]{\includegraphics[width=\columnwidth]{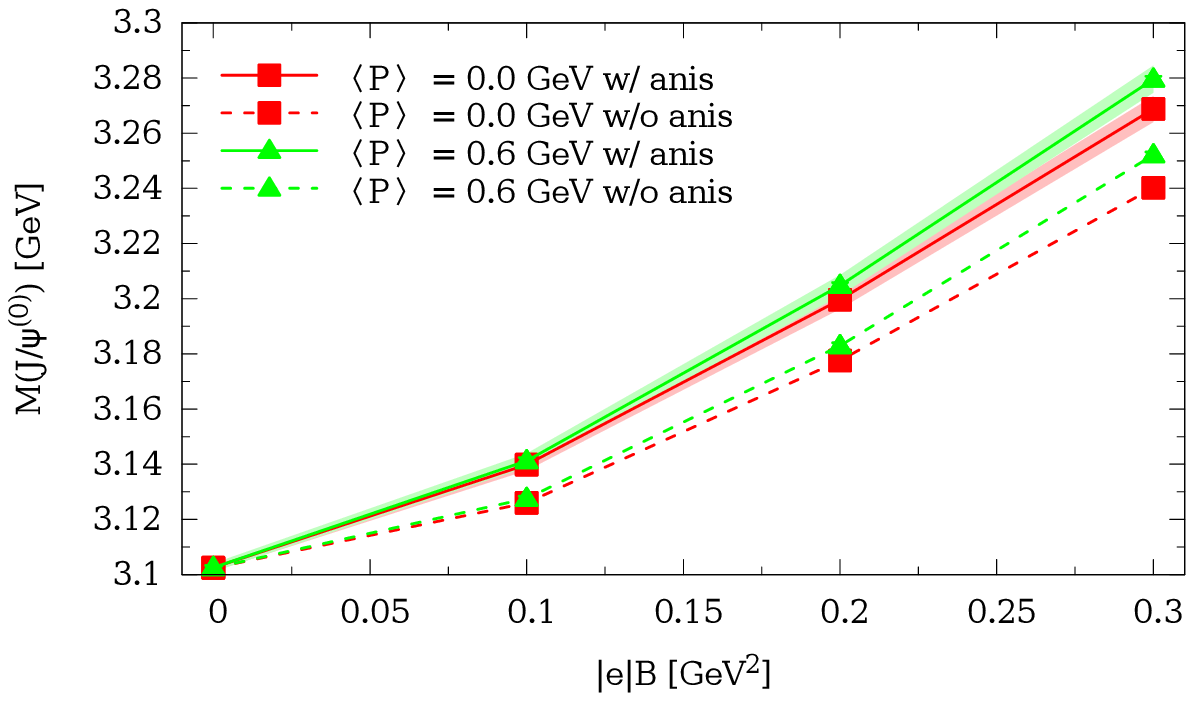}}
\caption{Behaviour of the masses of the charmonium $1S$ states with respect to
the magnetic field, both with and without the magnetic anisotropy in the
potential and for two different values of $\langle P\rangle$. Data points represent
the mass extracted making use of the central values of the parameters in
Tab.~\ref{table:anisoparam}, while shaded regions take into account
uncertainties on the parameters. (a) The $\eta_c$ singlet state. (b) The
$s_z=\pm~1$ components of the $J/\psi$ state. (c) The $s_z=0$ component of the
$J/\psi$ state.}
\label{fig:cc1s}
\end{figure}

We start by studying the $\eta_c$ and $J/\psi$ states, which correspond to the
$1S$ states of the charmonium at vanishing magnetic field.  The experimental
masses of these states are well reproduced by using the potential parameter
reported in Tab.~\ref{table:potparam} and their behaviour as a function of the
external magnetic field $B$ is shown in Fig.~\ref{fig:cc1s}. The spin
components $s_z=\pm 1$ of the $J/\psi$ stay degenerate also for $B\neq 0$ but
they split from the $s_z=0$ component, which mixes with the $\eta_c$ state.  

In order to isolate the contribution of the anisotropy in the potential, we
reported in Fig.~\ref{fig:cc1s} also the masses computed without taking into
account the anisotropy (i.e. with $\epsilon_1^{(\alpha)}=
\epsilon_1^{(\sigma)}=1$ and $\epsilon_2^{(\alpha)}=\epsilon_2^{(\sigma)}=0$).
It can be seen that in all cases the masses are increased by taking into account
the anisotropy; as a consequence, for the $\eta_c$ the dependence of the mass on
$B$ gets milder, while the opposite effect is observed for the $J/\psi$ states.
We note that for the $\eta_c$ and the $s_z=\pm 1$ components of the
$J/\psi$ the effect induced by the anisotropy is of the same order of magnitude
as the effect expected when no anisotropy is present. 

A non-vanishing kinetic momentum has an effect analogous to that of the
anisotropy: it reduces the dependence on $B$ of the $\eta_c$ mass while it
increases that of the $J/\psi$ states.  This effect is however quite small for
the values of the kinetic momenta explored and it is almost of the same 
magnitude of the uncertainties associated with the anisotropies. It cannot be
excluded that this effect increases for larger momenta, but to systematically
explore this regime a fully relativistic treatment would be required.

\begin{figure}[htb!]
\includegraphics[width=\columnwidth]{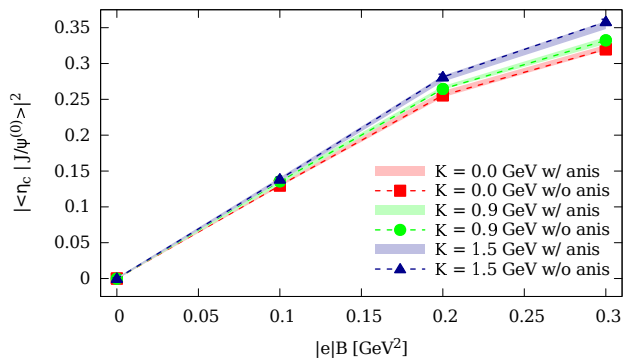}
\caption{Square modulus of the mixing between the $s_z=0$ component of the
$J/\psi$ state and the $\eta_c$ state, both with and without the magnetic anisotropy
in the potential and for several values of the pseudomomentum $\mbf{K}$.}
\label{fig:cc1smix}
\end{figure}

As previously noted, the magnetic field introduces a mixing between the
$\eta_c$ state and the $s_z=0$ component of the $J/\psi$ state.  The behaviour
of this mixing with respect to the magnetic field and the pseudomomentum $K$ is
shown in Fig.~\ref{fig:cc1smix}: while the dependence on the magnetic field is
quite strong, the mixing turn out to be almost independent of the value of the
pseudomentum.  Moreover, contrary to what happens for the masses, the
mixing is almost insensitive to the anisotropy of the potential.

\begin{figure}[htb!]
\subfigure{\includegraphics[width=\columnwidth]{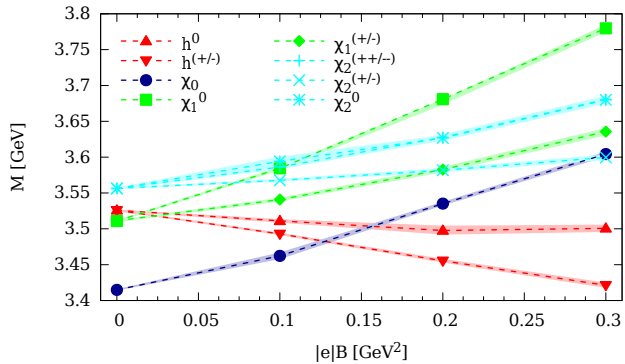}}
\caption{Behaviour of the masses of the charmonium $1P$ excited states with
respect to the magnetic field.  We report for simplicity only data obtained by
using the anisotropic form of the potential, the general features of this
picture are present also in the standard case of the central potential.}
\label{fig:cc1p}
\end{figure}

We now present some preliminary data for the $1P$ states. In this case the fine
spectrum of known $1P$ charmonium levels is not correctly reproduced for $B=0$,
a fact that is likely due to the absence of the spin-orbit coupling and other
relativistic corrections in the used Hamilonian. If we assume that these
relativistic terms do not depend on the external magnetic field, the computed
variation of the mass $\Delta m(B)$ due to a nonvanishing $B$ can be
used to shift the known $B=0$ values of the masses to obtain the spectrum at
$B\neq 0$. It is clear that this is an approximate procedure and that a more
precise study will be required to quantify the systematic error
introduced in this way; nevertheless this method is expected to give reliable
informations in the limit of large quark mass $m$, 
since relativistic effects are
suppressed by inverse powers of $m$. The results obtained in this way
are shown in Fig.~\ref{fig:cc1p} and the main difference with respect to the
$1S$ case is that in the $1P$ case several level crossings happen by increasing
the magnetic field. From Figs.~\ref{fig:cc1p} and \ref{fig:cc1s} it can be seen
that the gap between the $h$ states and the $J/\psi$ states 
is strongly reduced by increasing the magnetic field, a fact that may have
significant consequences on branching ratios of the $h$ meson.

\subsection{Bottomonium states}

\begin{figure}[htb!]
\subfigure[]{\includegraphics[width=\columnwidth]{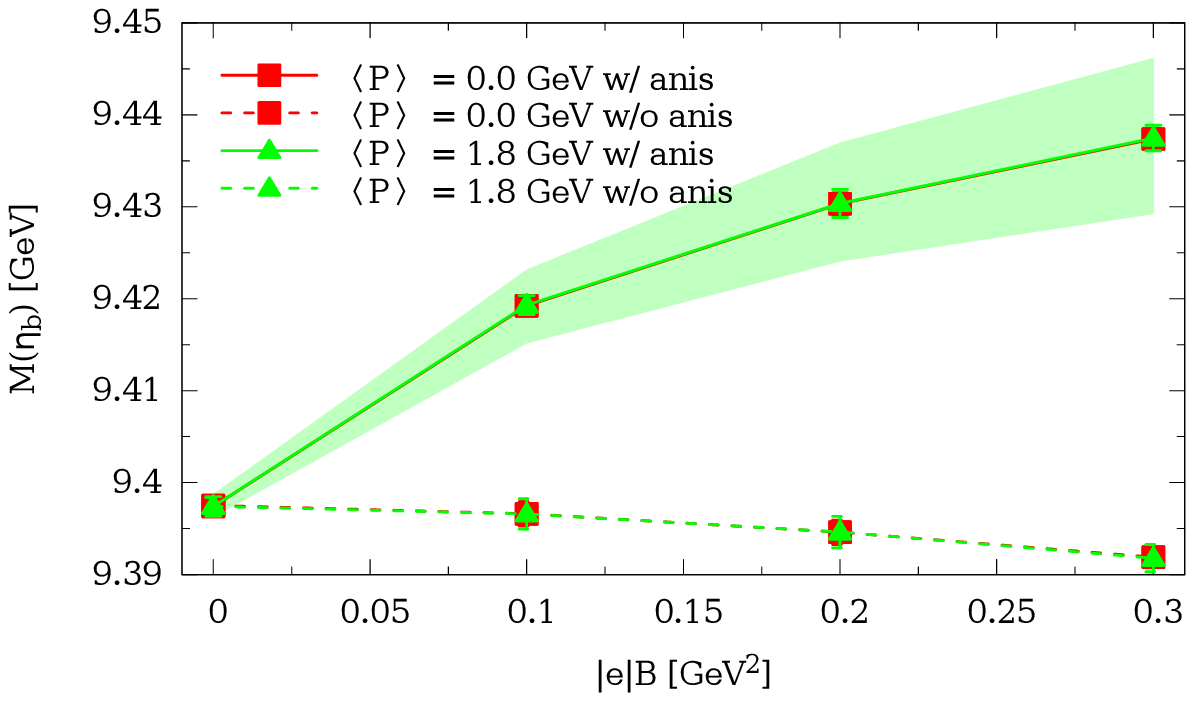}}
\subfigure[]{\includegraphics[width=\columnwidth]{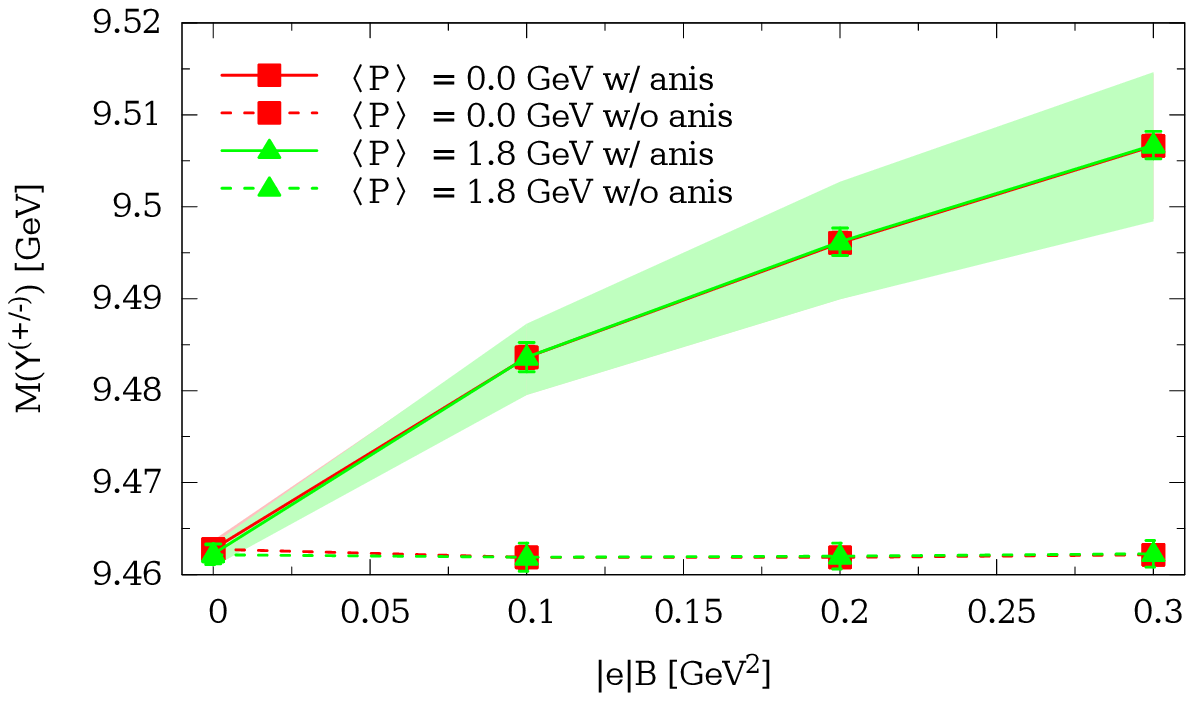}}
\subfigure[]{\includegraphics[width=\columnwidth]{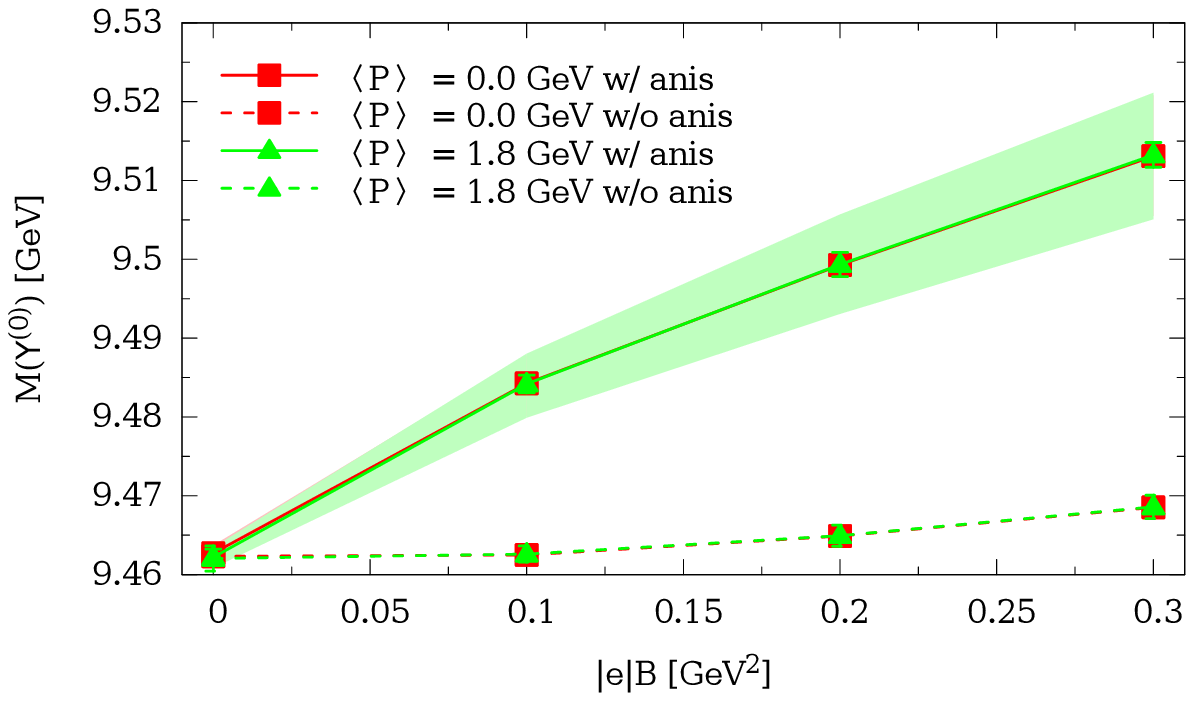}}
\caption{Behaviour of the masses of the bottomonium $1S$ states with respect to
the magnetic field, both with and without the magnetic anisotropy in the
potential and for several values of $\langle P\rangle$.
(a) The $\eta_b$ singlet state. (b) The $s_z=\pm 1$ components of the $\Upsilon$
state. (c) The $s_z=0$ component $\Upsilon$ state.}
\label{fig:bb1s}
\end{figure}

The effect of the anisotropy in the potential on bottomonium states is
qualitatively similar to that on charmonium states, however its quantitative
relevance is much larger, as can be seen from the spectrum of the $1S$ states
in Fig.~\ref{fig:bb1s}: the mass shift due to the anisotropy is about one order
of magnitude larger than the one due to quark magnetic momenta only. 
This is not unexpected, since the effect of the anisotropy in the potential is
independent of the quark masses, while quark magnetic momenta go to zero like $1/m$.
As a consequence the relative effect of the
anisotropy is much stronger for bottomonium than for charmonium.  

\begin{figure}[htb!]
\includegraphics[width=\columnwidth]{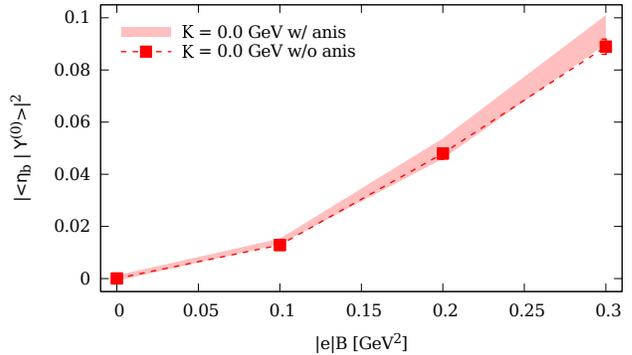}
\caption{Square modulus of the mixing between the $s_z=0$ component of the
$\Upsilon$ state and the $\eta_b$ state, both with and without the magnetic anisotropy in the potential
and for several values of the pseudomomentum $\mbf{K}$.} \label{fig:bb1smix}
\end{figure}

In Fig.~\ref{fig:bb1smix} we show the behaviour of the mixing between the
singlet state $\eta_b$ and the $s_z=0$ component of the $\Upsilon$ as a
function of the magnetic field. Although the value of the mixing is reduced by about 
a factor $3$ ($\approx m_b/m_c$) with respect to the charmonium case, its value is still 
largely independent on the anisotropy.

\begin{figure}[htb!]
\subfigure{\includegraphics[width=\columnwidth]{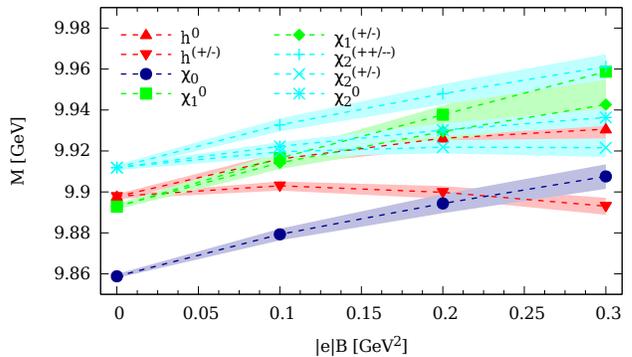}}
\caption{Behaviour of the masses of the bottomoniumm $1P$ excited states with
respect to the magnetic field.} \label{fig:bb1p}
\end{figure}

A preliminary analysis of the $1P$ states is shown in Fig.~\ref{fig:bb1p}: it
is obtained by the same procedure adopted for the charmonium, which for
bottomonium is expected to be more reliable. Also for bottomonium some level
crossings can be seen, however in this case the uncertainties, due to the
propagation of the error on the anisotropy, become quite large with respect to
the typical level separations.

\section{Summary and conclusions}
\label{summary}

The main purpose of this work has been the study of the possible influence of
strong external magnetic fields on the heavy-meson mass spectrum.  Following
the approach of Ref.~\cite{Alford}, we  used a non-relativistic model
to extract the lowest lying $c\bar{c}$ and $b\bar{b}$ states and the
magnetic field induced mixings among them.  As a new ingredient, we considered
the anisotropies induced by the magnetic field at a non-perturbative level in
the static quark-antiquark potential, as determined by a recent LQCD
investigation~\cite{anisostring}.  Moreover, we extended our analysis to the
first excited $1P$ states.

Various approximations are involved in our computation: relativistic
corrections were only partially included and any possible dependence of their
magnitude on $B$ was neglected.  Moreover, the exact angular dependence of the
static potential was obtained by a suitable interpolation of data reported in
Ref.~\cite{anisostring}.

A first issue of phenomenological interest regards the mixing between different
states, since that should lead to a modification of both decay patterns and
production rates. For instance, results suggest a contamination of the
semi-leptonic decay channels between the $\eta_c$ and the $J/\psi$ and between
the $\eta_b$ and $\Upsilon$ mesons. However, in this case such mixings are
observed even without taking into account the potential
anisotropy~\cite{Alford}, and we have verified that its inclusion does not lead
to significant quantitative changes.

On the other hand, results suggest that the presence of the anisotropy in the
potential has significant effects on the heavy-meson mass spectrum. An
increase of the masses is generally observed, with respect  to the case in
which the anisotropy is not included, the increase being of the order of
$30-40~\textrm{MeV}$ at the maximum magnetic field explored, $|e|B \simeq 0.3\
\textrm{GeV}^2$.  The effect is more dramatic for bottomonium states, where it
can even change, in some cases, the sign of the dependence of the mass on $B$:
the reason is that standard magnetic interaction terms in the Hamiltonian of
the system (see Eq.~\eqref{eq:redham}) are suppressed by the inverse of the
quark mass, while the non-perturbative effect on the static potential is mass
independent, hence it becomes dominant for bottomonium.

Finally, a new effect pointed out by our study regards the possible crossings
of $1P$ states as a function of $B$. Such crossings are observed even without
taking into account the anisotropy of the potential, however its presence 
makes them clearer. We stress that our results for $1P$ states are still preliminary:
in particular, regarding the level crossings, we assumed the experimentally
observed spectrum at $B = 0$, which can be reproduced for such states only by
taking into account relativistic corrections like the spin-orbit one. On the
other hand, the spin-orbit term was not taken into account to compute the
$B$-dependence of the spectrum, therefore a significant $B$-induced correction
to the this term could change the scenario.

One should consider that present results are obtained using the $T = 0$ form of the potential and assuming a
constant and uniform magnetic field.
The use of the $T=0$ potential is justified only for hard processes taking place in the initial 
stages of non-central collisions, before the strongly interacting medium thermalizes.
Future lattice simulations could provide information on magnetic field effects
at finite $T$, which could be relevant to production and decay rates of heavy
quark states in the thermalized medium; instead, it will be relatively more
difficult to take into account  the possible effects of inhomogeneities or time
dependence of the magnetic field distribution.  Further improvement on present
results could also be obtained once lattice simulations provide information
about the exact angular dependence of the static potential (which in the
present study was partially based on an ansatz) and, possibly, about the
spin-dependent part of the potential.

\section{Acknowledgement}

It is a pleasure to thank Michele Mesiti, Marco Mariti, Francesco Negro,
Francesco Sanfilippo and Michael Strickland for useful comments and
discussions.  Numerical simulations have been performed using resources
provided by the Scientific Computing Center at INFN-Pisa.

\appendix

\section{The numerical algorithm}\label{sec:algorithm} 

The algorithm used to numerically compute eigenvalues and eigenfunctions of
the reduced Hamiltonian \eqref{eq:redham} is the Finite Difference Time Domain
method (FDTD) described, e.g., in \cite{Sudiarta, Strickland}. The main idea of
this approach is the following: once the Schr\"{o}dinger equation is Wick
rotated to imaginary time
\begin{equation}\label{eq:euclschrod}
\left(\frac{\partial}{\partial\tau}+\hat{H}\right)\Psi(\mbf{r},\tau)=0\ , 
\end{equation}
the formal solution of the problem with initial condition
$\Psi(\mbf{r},0)=\zeta(\mbf{r})$ is
\begin{equation}\label{eq:euclschrodsol}
\Psi(\mbf{r},\tau) = \sum_{\alpha}\langle \Phi_{\alpha}|\zeta \rangle 
\Phi_{\alpha}(\mbf{r})e^{-\epsilon_{\alpha}\tau}\ ,
\end{equation}
where $\epsilon_{\alpha}$ and $\Phi_{\alpha}$ are the eigenvalues and
eigenfunctions of the Hamiltonian. By looking at the large time behaviour of
$\Psi(\mbf{r},\tau)$ we can thus identify the lowest $\epsilon_{\alpha}$ (and
the corresponding $\Phi_{\alpha}$) among those corresponding to states such
that $\langle \Phi_{\alpha}|\zeta\rangle\neq 0$.  By using an initial wave
function $\zeta(\mbf{r})$ with specific symmetries we can thus select the state
we are interested in and, in this work, hydrogen-like wave functions were used
to this purpose when $B=0$. States for $B\neq 0$ were extracted by
adiabatically switching on the magnetic field in an initial stage of the
evolution. This procedure turned out to work well for the low lying states,
however numerical instabilities emerge when applying this technique to higher
excited states.

From the numerical point of view the evolution is performed by introducing a
temporal and a spatial lattice spacing (denoted by $d\tau$ and $a$
respectively) and by approximating derivatives in Eq.~\eqref{eq:euclschrod}
with finite differences. In this way variables at time $\tau+d\tau$ can be
written in term of the ones at time $\tau$ (see \cite{Sudiarta} for details).

After every time evolution step we get an estimate of the bound state energy
and of other properties of the state we want to study; the process ends when
the variation of these quantities during a time interval $\tau_s$ goes below a
fixed threshold. The time interval $\tau_s$ was set to $1/M$, where $M$ is a
rough estimate of the mass for the state we are interested in. The precision of
$1\,\mathrm{MeV}$ was used as a stopping criterion for the energy determination
(this value can be used as the uncertainty in the energy computation). 

\begin{figure}[h]
\includegraphics[width=\columnwidth]{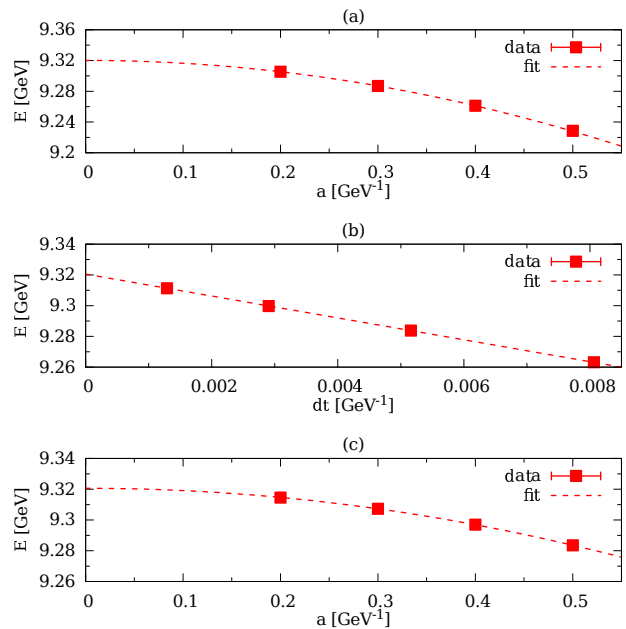}
\caption{Test of the continuum limit for parameters $\omega=1.5$ GeV,
$|e|B=0.9$ GeV$^{2}$ and $K=5.0$ GeV. Data correspond to the energy of the
eigenstate $\Phi_{(0,1)}^{(\pm)}$, while dashed lines are fit curves.  
(a) First $\mathrm{d}\tau \to 0$ and then $a\to 0$, the final continuum value
is $E_a=9.3205(13)\,\mathrm{GeV}$.
(b) First $a\to 0$ and then $\mathrm{d}\tau\to 0$,
$E_b=9.3206(13)\,\mathrm{GeV}$.
(c) $a\to 0$ with $a\propto d\tau^2$, $E_c=9.3201(12)\,\mathrm{GeV}$.} 
\label{fig:testcont}
\end{figure}

\begin{figure}[h]
\subfigure[]{\includegraphics[width=\columnwidth]{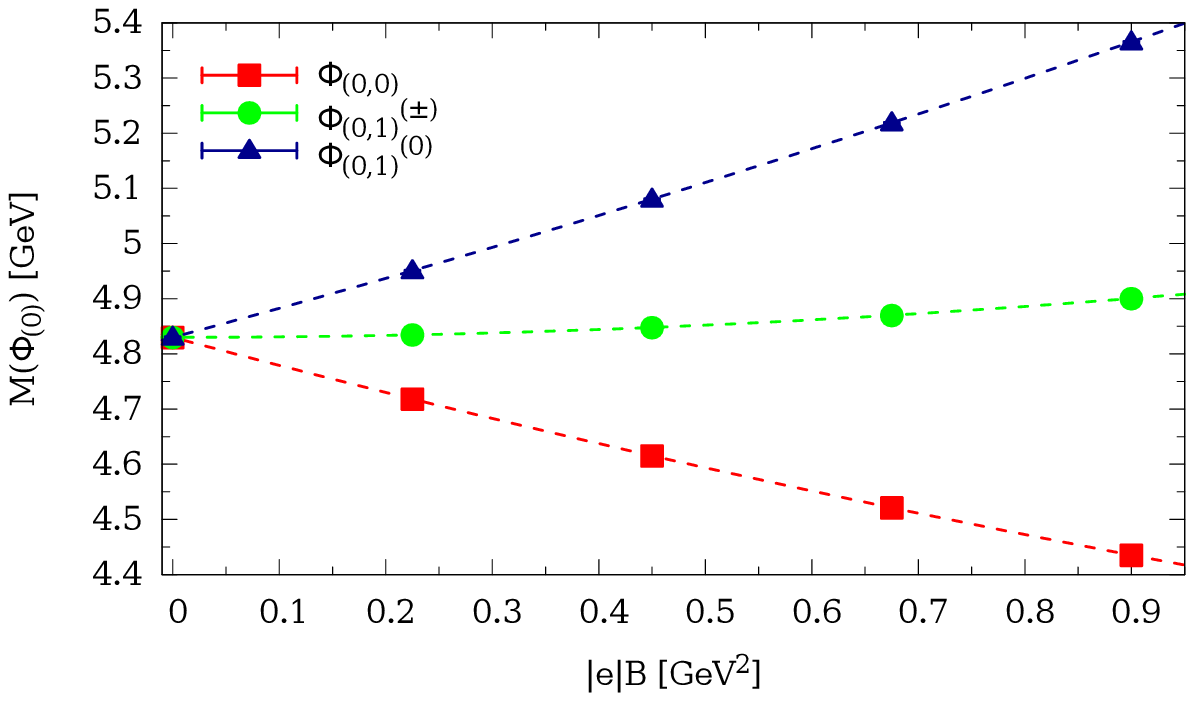}}
\subfigure[]{\includegraphics[width=\columnwidth]{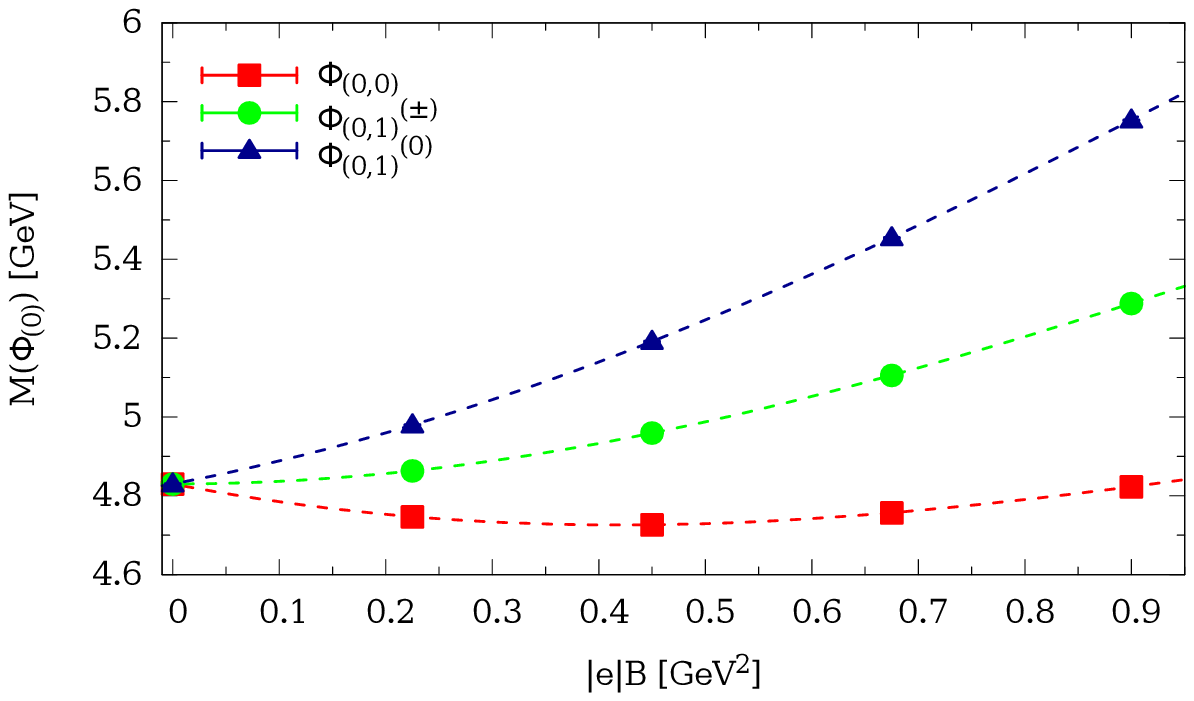}}
\caption{Behaviour of the masses for the eigenstates $\Phi_{(0,s)}$ with
respect to the magnetic field. Dots correspond to numerical data extrapolated
to the continuum, dashed curves represent the analytic expected behaviour. 
(a) Data corresponding to $|\mbf{K}|=0$. 
(b) Data corresponding to $|\mbf{K}|=5$ GeV, $\mbf{K}=|\mbf{K}|\mbf{\hat{x}}$.}
\label{fig:testmass}
\end{figure}

The numerical algorithm just described concerns the orbital part of the
Hamiltonian and can be applied, once the total spin is fixed, to the reduced
Hamiltonian Eq.~\eqref{eq:redham} (without the $\mbf{\mu}\cdot \mbf{B}$ terms)
together with the $V_{\sigma\sigma}$ potential defined by
Eq.~\eqref{eq:spinspin}. By using these eigenstates we can then evaluate the
matrix elements of the operator $(\mbf{\mu}_1+\mbf{\mu}_2)\cdot\mbf{B}$ and
diagonalize this matrix to extract the final eigenvalues and eigenstates.  

The operator $(\mbf{\mu}_1+\mbf{\mu}_2)\cdot\mbf{B}$ has non-vanishing matrix
elements only between the spin states $|00\rangle $ and $|10\rangle$ (see
Ref.~\cite{Alford} for details), however some care is needed here since, due to
the specific form of $V_{\sigma\sigma}$ adopted, nothing prevents a mixing
between states with different orbital momenta to occur (this problem is
relevant also for $B=0$). 
The effect of this mixing between different orbital momenta,
however, turns out to be rather small: in all the studied cases the overlap
$\langle \ell=0| \ell=1\rangle$ was compatible with zero within machine
precision and the overlap $\langle \ell=0| \ell=2\rangle $ was at most of the
percent level.

All the results presented in the main text where obtained by using a lattice of
physical spatial extent $V=(30~\textrm{GeV}^{-1})^3\simeq(6~\textrm{fm})^3$ and
several values of the lattice spacings, in order to extract the continuum
limit. The spatial lattice spacings used were $a=0.250, 0.375, 0.500,
0.625\,\textrm{GeV}^{-1}$ and the temporal lattice spacing was fixed by the
relation $\mathrm{d}\tau=ma^2/20$, $m$ being the quark mass.

\subsection{Test of the algorithm}

We report here the results of some tests performed to verify the correctness of
the algorithm implementation. As a testbed, we used the case of the harmonic
potential
\begin{equation} 
V(r) = \frac{1}{2}m\omega r^2\ ,
\end{equation}
which is analytically solvable (see Ref.~\cite{Herold, Alford}) and thus allows
for a direct check of the numerical data. Following the notation of
Ref.~\cite{Alford}, eigenstates are classified by the quantum numbers
$\mbf{K},n_{\perp},n_z,\ell,s$ and $s_z$, where $\mbf{K}$ is the
pseudo-momentum defined in Eq.~\eqref{eq:K}, $s$ is the total spin of the
system, $s_z$ its projection on the $z$-axis and $n_{\perp},n_z,\ell$ are
quantum numbers specific for the harmonic potential. For fixed spin $s$, the
ground state will be denoted by $\Phi_{(0,s)}$ and corresponds to the case
$n_{\perp}=n_z=\ell=0$ (the superscripts $\pm$ and $0$ will be used to denote
the $s_z$ components).

As a first test we verified that, for fixed physical parameters, the same
result is obtained by performing the continuum limit in different ways: it is
possible to first extract the limit for $a\to 0$ at fixed $\mathrm{d}\tau$ and
then send $\mathrm{d}\tau$ to zerol; alternatively,
 it is possible to exchange the order of the
limits or to perform both limits together. Since in our implementation
discretization errors are linear in $\mathrm{d}\tau$ and quadratic in $a$, it
is convenient to perform the limit by using a relation of the form
$\mathrm{d}\tau\propto a^2$ with a fixed proportionality factor.  

The three possible ways to continuum extrapolate are compared in
Fig.~\ref{fig:testcont} and they nicely agree with each other within errors.
In Fig.~\ref{fig:testmass} the final results (i.e. continuum extrapolated and
taking into account spin mixing) for the energies of the low lying states are
compared with the theoretical expectations, and in all cases a perfect
agreement is found.

\end{document}